\documentclass[twocolumn,superscriptaddress,showpacs]{revtex4}
\usepackage{amstext}
\usepackage{amsmath}            
\usepackage{amssymb}            
\usepackage{graphicx}           
\usepackage{latexsym}

\newtheorem{thm}{Theorem}

\newtheorem{prop}[thm]{Property}

\newcommand{\ket}[1]{\left\vert#1\right\rangle}
\newcommand{\bra}[1]{\left\langle#1\right\vert}
\newcommand{\sprod}[2]{\left\langle#1\right\vert\left.#2\right\rangle}

\newcommand{\Q}{\widetilde{Q}}

\begin{document}
\title{The role of auxiliary states in state discrimination with linear optical devices.}
\author{Angelo Carollo}
\author{G.Massimo Palma}
\affiliation{Dipartimento di Scienze Fisiche ed Astronomiche and INFM-Unit\`a di Palermo,\\ Via Archirafi 36,
I-90123 Palermo, Italy}

\date{\today}
\begin{abstract}
 The role of auxiliary photons in the problem of identifying a
 state secretly chosen from a given set of L-photon states is
 analyzed. It is shown that auxiliary photons do not increase the
 ability to discriminate such states by means of a
 global measurement using only optical linear elements,
 conditional transformation and auxiliary photons.
\end{abstract}
 \maketitle

\section{Introduction}
Linear quantum optical devices have proved to be ideal system for
the experimental implementation of several quantum information
processing protocols like quantum cryptography \cite{qcrypto},
quantum teleportation \cite{teleport} and quantum dense
coding~\cite{qdense} to mention some. One of the main limitation
of such systems is the difficulty to implement conditional
dynamics as photons interfere but hardly interact with each other.
However recently a proposal to implement probabilistic quantum
computation~\cite{milburn} making use of linear optical devices
has been put forward. In this proposal the difficulty to achieve
experimentally conditional dynamics is circumvented by making use
of auxiliary photons and obtaining a probability of success
asymptotically close to one. On the other hand the problem of
distinguishing completely photon states secretly chosen from a
given set by means of linear devices has been addressed in several
recent papers. In particular the set of Bell states has been
considered in~\cite{norbert,norbert2} while the set of "non local
without entanglement" \cite{bennett} states has been analyzed
in~\cite{nostro}.

In this paper we will analyze the role of auxiliary photons in the
problem of identifying the elements of an arbitrary  set of
orthogonal N-photon states by means of linear optical devices. We
will show that auxiliary photons cannot increase our ability to
identify which state has been chosen from a given input set.

\begin{figure}[ht]
\centering
\includegraphics[width=8cm]{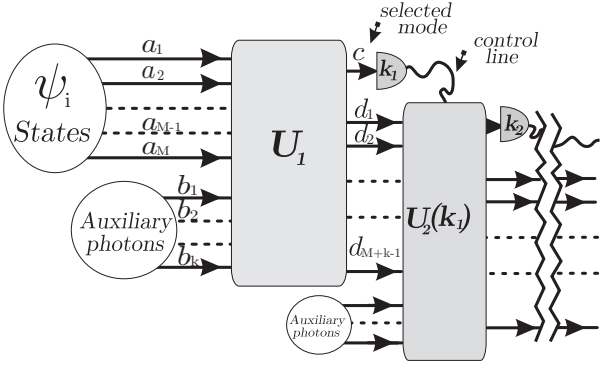}
\caption{Cascade setup in which the modes of the system states are
mixed in a first "box" with auxiliary modes. Selected output mode
is then measured and depending on its outcome the remaining output
modes are fed in a new box. The process can be repeated over and
over again} \label{cascade}
\end{figure}

The experimental setup we will consider is the one already
discussed in~\cite{norbert,nostro}. Suppose that we want to
discriminate {\em exactly} $L$-photon states over $M$ modes,
randomly chosen from a known set of $K$ states. In our ideal setup
the modes of the input states are mixed with an arbitrary number
of auxiliary modes in a "black box" consisting of optical linear
devices. The output modes of this box will be linked to the input
ones by a unitary transformation $U$. It has been shown that any
such unitary transformations of modes can be obtained by means of
linear optical devices~\cite{reck}, like beam splitters and phase
shifters. To ensure the largest possible generality in our
measurement apparatus we will assume the possibility to perform
conditional measurements. In practice this means what follows:
assume that a measurement is made on one selected output mode
while the others are kept in a delay loop  and that, according to
the outcome of the measurement, these modes are fed into a
selected further black box, in a cascade setup ( see
figure~\ref{cascade} ). The final assumption we will make is that
our detectors have the ability to distinguish the number of
incident photons. Although this assumption is unrealistic we have
made it in order to guarantee the largest possible generality.

A first strategy to implement a measurement could be to mix the
modes by means of linear devices and than perform, with the
previously described cascade setup, conditional measurement on the
output modes of such device. However, following \cite{norbert,
nostro} we will adopt a more general strategy. We will assume to
have at our disposal a set of as many additional modes as we like,
here indicated with bosonic creation operators $c^{\dagger}_j$,
with any number of photons we like and we will assume that these
auxiliary modes can be mixed with modes $a^{\dagger}_i,
b^{\dagger}_k$ in a black box.

In this scenario we will show that the use of auxiliary photons
does not help in increasing the distinguishability of the input
states. In other words we will demonstrate that given any two
input $L$-photon states, they are completely distinguishable in
the presence of auxiliary photons only if they are completely
distinguishable in the absence of auxiliary photons. We will show
that this is a consequence of the fact that, given  a measurement
outcome in the selected output mode, the error probability in
presence of auxiliary photons has a direct linear relation with
the error probability in absence of auxiliary photons. This
assures that auxiliary photons cannot improve complete
distinguishability.

The paper is structured as follows: in the next section we will
describe our measurement setup and in section III we will prove
our statement. As the mathematics involved is rather tedious to
follow some of the details are discussed in appendix.

\section{Statement of the problem}
As described already our measuring apparatus consists of a cascade
of "black boxes", in which modes are linearly mixed, and partial
measurements, which determine the sequence of unitary mixing. The
first of such black box, denoted by  $U_1$, is made out of linear
optical elements and its input and output are a set of bosonic
modes. The joint input modes consist of M "system" modes
$\hat{a}^{\dagger}_i$ and an arbitrary number of auxiliary modes
$\hat{b}^{\dagger}_i$. These input modes are unitarily mixed in
the box into a set of output modes $\hat{c}^{\dagger},
\hat{d}^{\dagger}_i$ where the $\hat{c}^{\dagger}$ mode is the one
on which a measurement will be performed. The measurement outcome
determines the specific unitary mixing that will be performed in
next step of the measurement, consisting of a second box $U_2$.
While the measurement on mode $\hat{c}^{\dagger}$ is performed the
photons in the remaining $\hat{d}^{\dagger}_i$ modes are kept in a
waiting loop. The whole measurement procedure consists of a
cascade of conditional measurements as described above.

The input state of the first block $U_1$ of the measuring
apparatus is of the form $\ket{\psi_i^{tot}} =
\ket{\psi_{aux}}\otimes\ket{\psi_i}$, where $\ket{\psi_i}$ is the
state randomly chosen from a set of $K$ $L$-photon states of $M$
modes we want to identify and $\ket{\psi_{aux}}$ is the state of
the auxiliary modes. Such input state $\ket{\psi_i^{tot}}$ can be
written as

\begin{equation}
\ket{\psi_i^{tot}} = \ket{\psi_{aux}}\otimes\ket{\psi_i} =
P_{aux}(\hat{b}^{\dagger}_k) P_i(\hat{a}^{\dagger}_n)\ket{0}
\label{input}
\end{equation}

where $P_i(\hat{a}^{\dagger})$ is a polynomial of degree $N$ and
$P_{aux}(\hat{b}^{\dagger}_k)$ is a polynomial of arbitrary degree
in the $\hat{b}^{\dagger}_k$

The corresponding output state is

\begin{equation}
\ket{\psi^{tot}_i}=
\tilde{P}_{aux}(\hat{c}^{\dagger},\hat{d}^{\dagger}_k)\tilde{P}_{\psi_i}(\hat{c}^{\dagger},\hat{d}^{\dagger}_k)\ket{0}
\label{output}
\end{equation}

Where $\tilde{P}_{aux}(\hat{c}^{\dagger},\hat{d}^{\dagger}_k)$ and
$\tilde{P}_{\psi_i}(\hat{c}^{\dagger},\hat{d}^{\dagger}_k)$ are
nothing but $P_{aux}(\hat{b}^{\dagger}_k)$ e
$P_i(a^{\dagger}_k)\ket{0}$ written in terms of the creation and
annihilation operators at the output of $U_1$.

We can expand ${\tilde P}_{aux}, {\tilde P}_{\psi_i}$ in terms of
decreasing powers of $\hat{c}^{\dagger}$ as follows

\begin{align}
\tilde{P}_{aux}(\hat{c}^{\dagger},\hat{d}^{\dagger}_k)&=\label{espanaux}
\sum_{n=0}^{n_a}(\hat{c}^{\dagger})^n\Q_a^{(n)}(\hat{d}^{\dagger}_k)\\
\tilde{P}_{\psi_i}(\hat{c}^{\dagger},\hat{d}^{\dagger}_k)&=
\sum_{n=0}^{n_s}(\hat{c}^{\dagger})^n\Q_{\psi_i}^{(n)}(\hat{d}^{\dagger}_k)\label{espanpsi}
\end{align}

In (\ref{espanpsi}) $n_s$ is the largest order in $\hat{c}^{\dagger}$ for the ${\tilde P}_{\psi_i}$ with $i=1..K$
and by definition is independent on index $i$ (${\tilde Q}_{\psi_i}$ can be zero for some $i$). Analogously $n_a$
is defined as the order in $\hat{c}^{\dagger }$ of polynomial  ${\tilde P}_{aux}$. We can therefore rewrite
(\ref{output}) as

\begin{equation}
\ket{\psi^{tot}_i}=\sum_{n,m=0}^{n_a,n_s}(\hat{c}^\dag)^{n+m}\Q_a^{(n)}(\hat{d}^\dag_n)
\Q_{\psi_i}^{(m)}(\hat{d}^\dag_k)\ket{0}
\end{equation}

Let's suppose now that the number of photons on the selected mode
$\hat{c}$ is measured. If $N$ is the outcome of such measurement
the (unormalised) conditional state of the remaining modes can be
we written as

\begin{equation}\label{stcond}
\ket{\psi^{N}_i}=\sum_{k}\Q_a^{(k)}\Q_{\psi_i}^{(N-k)}\ket{0}
\end{equation}

where $\max\{0,N-n_s\}\leq k\leq\min\{n_a,N\}$.

If the input states are to be distinguishable the conditional states $\ket{\psi^{N}_i}$ \emph{must be orthogonal
for each possible value of $N$}, i.e.

\begin{equation}\label{dist}
\sprod{\psi_i^{N}}{\psi_j^{N}} = 0 \hspace{0.5cm} \forall N, i\neq
j
\end{equation}

Of particular interest in the following will be the case in which
no photon is present in the auxiliary modes. In this case the
distinguishability condition (\ref{dist}) becomes:

\begin{equation}\label{distwithout}
  \sprod{\psi_i^{N}}{\psi_j^{N}}_{n_a=0}\!\! \propto \bra{0}
\Q_{\psi_i}^{(N)\dag}\Q_{\psi_j}^{(N)}\ket{0} = 0
\hspace{0.5cm}\forall N, i\neq j
\end{equation}

where $\ket{\psi_i^N}_{n_a=0}$ is the conditional output state obtained from $\psi_i$ in the absence of auxiliary
photons when $N$ photons are measured in mode $\hat{c}$. Here we have made of use of expression (\ref{stcond}).

Out of the possible outcomes of the measurement of the number $N$
of photons in mode $\hat{c}$ we will concentrate on some
particular outcomes, namely those for which
\begin{equation}\label{higestnumber}
n_a\leq N\leq n_a+n_s
\end{equation}

The reason of this particular choice will be shortly evident. Now,
we define the following $n_s+1$-dimensional vectors:

\begin{equation}
\mathbb{V}= \left(\begin{array}{c}
\sprod{\psi_i^{n_a+n_s}}{\psi_j^{n_a+n_s}}\\
\vdots\\
\sprod{\psi_i^{n_a+1}}{\psi_j^{n_a+1}}\\
\sprod{\psi_i^{n_a}}{\psi_j^{n_a}}
\end{array}\right)
\quad \mathbb{U}= \left(\begin{array}{c}
\sprod{\psi_i^{n_s}}{\psi_j^{n_s}}_{n_a=0}\\
\vdots\\
\sprod{\psi_i^{1}}{\psi_j^{1}}_{n_a=0}\\
\sprod{\psi_i^{0}}{\psi_j^{0}}_{n_a=0}
\end{array}\right)
\label{vectorV}
\end{equation}

whose elements are scalar products of conditional states after the
measurement of $n$ photons in mode $\hat{c}$, with and without the
auxiliary photons respectively. Using this vectors, conditions
(\ref{distwithout}) can be rephrased as:
\begin{subequations}
\begin{equation}\label{vector:cond:without}
  \mathbb{U}=\underline{\mathbf{0}}
\end{equation}
while a necessary condition for distinguishability with auxiliary
photon can be expressed as  (see (\ref{dist})):
\begin{equation}\label{vector:cond:with}
  \mathbb{V}=\underline{\mathbf{0}}
\end{equation}

\end{subequations}

The central point of this paper is that, as we will show, vectors
$\mathbb{V}$ and $\mathbb{U}$ are linearly connected through a
matrix whose determinant is not vanishing,i.e.
\begin{equation}\label{linear}
  \mathbb{V}=\mathbf{M}\mathbb{U}\qquad \det(\mathbf{M})\ne 0
\end{equation}
The elements of this matrix only depend on the auxiliary photon
states and are indedependent on the states of the system. This
implies that conditions (\ref{vector:cond:without}) and
(\ref{vector:cond:with}) are completely equivalent or, in other
words, that \emph{distinguishability in absence of auxiliary
photon is a necessary condition for distinguishability in presence
of auxiliary photons.}

\section{Proof that auxiliary photons do not increase complete distiguishability}
\label{nogo} In order to demonstrate the previous statement we
define the following $n_s+1$-dimensional vector:
\begin{equation}\label{uprime}
  \mathbb{U}'\equiv\left(\begin{array}{c}
\bra{0}\Q_{\psi_i}^{(n_s)\dag}\Q_{\psi_j}^{(n_s)}\ket{0}\\
\vdots\\
\bra{0}\Q_{\psi_i}^{(1)\dag}\Q_{\psi_j}^{(1)}\ket{0}\\
\bra{0}\Q_{\psi_i}^{(0)\dag}\Q_{\psi_j}^{(0)}\ket{0}
\end{array}\right)
\end{equation}
From eq.~(\ref{distwithout}) immediately follows that
$\mathbb{U}'$ is linearly connected with $\mathbb{U}$.

In this section we will demonstrate the following

{\bf Theorem} {\it Given the vectors $\mathbb{V}$ and
$\mathbb{U}'$, defined in eq~(\ref{vectorV}) and
eq.~(\ref{uprime}) respectively, it follows that:

\begin{equation}\label{linear1}
  \mathbb{V}=\mathbf{M}'\mathbb{U}'
\end{equation}
where $\mathbf{M}'$ is a triangular matrix of this form:

\begin{equation}
\mathbf{M}'=
 \left(
  \begin{array}{ccccc}
    {\cal D}     & 0                         & \cdots &      0                 &      0\\
    m_{1\, 0}      & {\cal D}                  & \cdots &      0                 &      0\\
    m_{2\, 0}      &m_{2 1}         & \ddots &      0                 &      0\\
    \vdots                  &\vdots                     &  &\vdots                  &\vdots \\
    m_{n_s\, 0}    &m_{n_s 1}       & \cdots &m_{n_s\; n_s-1}&{\cal D}
 \end{array}
 \right)
\end{equation}
with
\begin{equation}\label{diag}
  {\cal D}\equiv\bra{0}\Q_a^{(n_a)\dag}\Q_a^{(n_a)}\ket{0}.
\end{equation}
}

{\bf Remark}  {\it By definition ${\cal
D}=\|\Q_a^{(n_a)}(\hat{d}^{\dagger}_k)\ket{0}\|^2>0$, being
$\Q_a^{(n_a)}$ the first non vanishing term of the
expansion~(\ref{espanaux}).From this follows that

\begin{equation}\label{detm1}
  \det\left(\mathbf{M}'\right)={\cal D}^{n_s+1}>0.
\end{equation}
which implies eq.~(\ref{linear}). }

Before entering into the details of our proof  we briefly
introduce some notation.

 From~(\ref{stcond}) follows that the scalar product between the
(unormalised) states $\ket{\psi^{N}_i}, \ket{\psi^{N}_j}$ obtained
after the measurement of $N$ photons in mode $\hat{c}$ is
\begin{equation}\label{prodN}
\sprod{\psi^N_i}{\psi^N_j}= \sum_{n,m}
\bra{0}\Q_{\psi_i}^{(N-m)\dag} \Q_a^{(m)\dag} \Q_a^{(n)}
\Q_{\psi_i}^{(N-n)}\ket{0}
\end{equation}

with $max\{0,N-n_s\}\leq n,m \leq min\{n_a,N\}$.

Let's define $N_m=n_a+n_s$ and express the eq. (\ref{prodN}) in an
alternative form:
\begin{equation}\label{prodN1}
\sprod{\psi^{N_m-s}_i}{\psi^{N_m-s}_j}=\sum_{n,m}{\cal
C}_{m,n}^{(s)}(i,j)
\end{equation}
where $s$ is defined as: $s\equiv N_m-N$, and obviously $0\leq s\leq N_m$, while ${\cal C}_{n,m}^{(s)}$ is defined
as:
\begin{multline}\label{cij}
\!{\cal C}_{n,m}^{(s)}(i,j)\equiv \\\equiv\bra{0}\Q_{\psi_i}^{(n_s-n)\dag} \Q_a^{(n_a-s+n)\dag} \Q_a^{(n_a-s+m)}
\Q_{\psi_j}^{(n_s-m)}\!\!\ket{0}
\end{multline}

In this section we will make use of the following properties of
coefficients ${\cal C}_{n,m}^{(s)}(i,j)$, which are demonstrated
in appendix \ref{appA}:
\begin{prop}
Symmetry:
\begin{equation*}
  {\cal C}_{n,m}^{(s)}(i,j)={\cal C}_{m,n}^{(s)}(i,j)
\end{equation*}
\end{prop}

\begin{prop}
Recurrence relation:
\begin{multline}\label{recurrence}
{\cal C}_{m,n}^{(s)}(i,j)=\\\delta_{n,m}\!\!\bra{0}\Q_{\psi_i}^{(n_s-n)\dag}\Q_{\psi_j}^{(n_s-n)}\ket{0}\!\!\bra{0}
\Q_a^{(n_a-s+n)\dag} \Q_a^{(n_a-s+n)}\ket{0}-\\-\sum_{k}^ k\binom{n_a-s+m+k}{k}\!\!\binom{n_s-n+k}{k} {\cal
C}_{n-k,m}^{(s-k)}(i,j)
\end{multline}
with $n\geq m$,  $1 \leq k \leq \min\{n,s-m\}$, and where
$\delta_{n,m}$ is the Kroneker symbol.
\end{prop}

It is straightforward to verify by recursion that Property 2
implies the following expression for the coefficients ${\cal
C}_{m,n}^{(s)}(i,j)$:

\begin{subequations}\label{As}
\begin{equation}
  {\cal
C}_{m,n}^{(s)}(i,j)=\sum_p{\cal
A}_p^{(s)}(n,m)\bra{0}\Q_{\psi_i}^{(n_s-p)\dag}\Q_{\psi_j}^{(n_s-p)}\ket{0}
\end{equation}
where \hbox{$\min\{n,m\}\leq p \leq \max\{0,n+m-s\}$}, and:
\begin{align}
&{\cal A}_n^{(s)}(n,n)=\bra{0}\Q_a^{(n_a-s+n)\dag}\Q_a^{(n_a-s+n)}\ket{0}\\\label{As:c} &{\cal
A}_p^{(s)}(n,m)=\\&\!-\!\!\sum_{k} k!\!\binom{n_a-s+m+k}{k}\!\!\binom{n_s-n+k}{k} {\cal
A}_{p}^{(s-k)}(n-k,m)\nonumber
\end{align}

where the eq.~(\ref{As:c}) is valid for $n\geq m$, and $1 \leq k \leq \min\{n-p,s-m\}$.
\end{subequations}

The above expression gives us a recursive method for calculating
all the coefficient of expression (\ref{prodN1}) once the
auxiliary photons states are known. With the help of
eqs.~(\ref{As}) it is possible to derive the following properties
of coefficients ${\cal A}_{p}^{(s)}(n,m)$,

\begin{itemize}
  \item [-] ${\cal
A}$'s only depend on the auxiliary states and do not depend at all
on the states $\psi_i$ and $\psi_j$.
  \item [-] ${\cal
A}$'s are real numbers.
  \item [-] ${\cal
A}$'s are symmetric in respect to the indices n and m.
\end{itemize}

It is now possible to cast equation (\ref{prodN1})  as a sum whose
elements factor in a product of coefficients depending only on
system states with coefficients  depending only on auxiliary
states. Let's consider eq. (\ref{prodN1}) in the case of $s\leq
n_s$:
\begin{multline}\label{prodN2}
\sprod{\psi^{N_m-s}_i}{\psi^{N_m-s}_j}=\sum_{n,m=\max\{0,s-n_a\}}^s{\cal
C}_{m,n}^{(s)}(i,j)\\
={\cal C}_{s,s}^{(s)}(i,j)+{\sum_{n,m}}'{\cal C}_{m,n}^{(s)}(i,j)
\end{multline}
where the sum ${\sum'}_{n,m}$ is extended over
\hbox{$\max\{0,s-n_a\}\leq n,m\leq s\quad$} with $\qquad
min\{n,m\}<s$. From eqs.~(\ref{As}) follows that

\begin{equation}\label{coeffD}
  {\cal C}_{s,s}^{(s)}(i,j)={\cal D}\cdot\bra{0}\Q_{\psi_i}^{(n_s-s)\dag}\Q_{\psi_j}^{(n_s-s)}\ket{0}
\end{equation}
where $\mathcal{D}$, the coefficient previously defined as
\begin{equation}\label{coeffD1}
  {\cal D}=\bra{0}\Q_a^{(n_a)\dag}\Q_a^{(n_a)}\ket{0}={\cal
A}_{s}^{(s)}(s,s)
\end{equation}
is by definition independent from $s$ ; while in the second term
of (\ref{prodN2})
\begin{equation}\label{other:coeff}
  {\cal C}_{m,n}^{(s)}(i,j)=\sum_p{\cal
A}_p^{(s)}(n,m)\bra{0}\Q_{\psi_i}^{(n_s-p)\dag}\Q_{\psi_j}^{(n_s-p)}\ket{0}
\end{equation}

where \hbox{$\max\{n+m-s\}\leq p\leq\min\{n,m\}<s$}. Finally with the help of the
eqs.~(\ref{coeffD},~\ref{coeffD1},~\ref{other:coeff})
we can write:

\begin{multline*}
\sprod{\psi^{N_m-s}_i}{\psi^{N_m-s}_j}={\cal
D}\cdot\bra{0}\Q_{\psi_i}^{(n_s-s)\dag}\Q_{\psi_j}^{(n_s-s)}\ket{0}+\\+ \sum_{p}{\cal
B}_{p}^{(s)}\cdot\bra{0}\Q_{\psi_i}^{(n_s-p)\dag}\Q_{\psi_j}^{(n_s-p)}\ket{0}
\end{multline*}

with \ $\ \max\{0,s-2n_a\}\leq p\leq s-1\ $\  and

\begin{equation*}
{\cal B}_{p}^{(s)}=\sum_{n,m}{\cal A}_{p}^{(s)}(n,m)
\end{equation*}

This expression allows us to write:
\begin{equation}\label{linear2}
  \mathbb{V}=\mathbf{M}'\mathbb{U}'
\end{equation}
where

\begin{equation}
\mathbf{M}'=
 \left(
  \begin{array}{ccccc}
    {\cal D}                & 0                         & \cdots &      0                 &      0\\
    {\cal B}_{0}^{(1)}      & {\cal D}                  & \cdots &      0                 &      0\\
    {\cal B}_{0}^{(2)}      &{\cal B}_{1}^{(2)}         & \cdots &      0                 &      0\\
    \vdots                  &\vdots                     & \ddots &\vdots                  &\vdots \\
    {\cal B}_{0}^{(n_s-1)}  &{\cal B}_{1}^{(n_s-1)}     & \cdots &{\cal D}                &      0\\
    {\cal B}_{0}^{(n_s)}    &{\cal B}_{1}^{(n_s)}       & \cdots &{\cal B}_{n_s-1}^{(n_s)}&{\cal D}
 \end{array}
 \right)
\end{equation}
whose determinant is
\begin{equation}\label{detm2}
  \det\left(\mathbf{M}'\right)={\cal D}^{n_s+1}>0.
\end{equation}

This completes our proof.

\section{conclusions}

In this manuscript we have discussed the role of auxiliary photons
in state discrimination with linear optical devices. The cascade
setup we have considered is of large generality.We have shown that
for such setup auxiliary photons do not increase {\em complete}
distinguishibility. This results wants to be a contribution to the
assessment of the role of resources in quantum information
processing with linear optical devices.


\section*{Acknowledgments}
We would like to thank  J.Calsamiglia,  N.L\"utkenhaus, C.Simon
and A.Zeilinger for helpful discussions. This work was supported
in part by the EU under grants TMR - ERB FMR XCT 96-0087 - "The
Physics of Quantum Information" IST - 1999 - 11053 -
EQUIP,"Entanglement in Quantum Information Processing and
Communication".


\appendix

\section{}\label{appA}
In this section we will prove Property 1 and Property 2 of
section~\ref{nogo}. To this goal we will make use of the following
lemma:

{\bf Lemma} {\it If
$\left[\tilde{P}_{aux}\tilde{P}_{\psi_i}\right]=0$, where
$\tilde{P}_{aux}$ and $\tilde{P}_{\psi_i}$ are defined by
(\ref{output}), then:
\begin{equation}\label{recurs}
 \Q_{a}^{(n)}\Q_{\psi_i}^{(m)\dag}=\sum_{k}
k!\binom{m+k}{k}\!\!\binom{n+k}{k}
\Q_{\psi_i}^{(m+k)\dag}\Q_{a}^{(n+k)}
\end{equation}
where $ 0 \leq k \leq\min\{n_s-m,n_a-n\}$. }

\emph{Proof}. From eqs.~(\ref{espanaux})
and~(\ref{espanpsi})follows:
\begin{align}
  &\left[\tilde{P}_{aux}\tilde{P}_{\psi_i}\right]=\label{commut}
  \\&=\sum_{n=0}^{n_a}\sum_{m=0}^{n_s}
  \left\{\hat{c}^{\dagger n}\hat{c}^m \Q_{a}^{(n)}\Q_{\psi_i}^{(m)\dag}-
  \hat{c}^m\hat{c}^{\dagger n} \Q_{\psi_i}^{(m)\dag}\Q_{a}^{(n)}
  \right\}=0\nonumber
\end{align}

Furthermore  the following property for creation and annihilation
operators holds:

\begin{equation}\label{cc}
  \hat{c}^m\hat{c}^{\dagger n}=\sum^{\min\{n,m\}}_{k=0}
  k!\binom{m}{k}\!\!\binom{n}{k}
  \hat{c}^{\dagger
n-k}\hat{c}^{m-k}
\end{equation}

Inserting eq.(\ref{cc}) in eq.(\ref{commut}) and reordering
elements in the sum it can be shown that

\begin{equation}\label{sommacommut}
  \left[\tilde{P}_{aux}\tilde{P}_{\psi_i}\right]=
  \sum_{n=0}^{n_a}\sum_{m=0}^{n_s}\hat{f}_{n,m}(\hat{d}^{\dagger}_k,\hat{d}^m_k)\hat{c}^m\hat{c}^{\dagger n}=0
\end{equation}

where

\begin{align}
&\hat{f}_{n,m}=\left\{\Q_{a}^{(n)}\Q_{\psi_i}^{(m)\dag}-\right. \nonumber\\&\left.-\sum_{k}
k!\binom{m+k}{k}\!\!\binom{n+k}{k} \Q_{\psi_i}^{(m+k)\dag}\Q_{a}^{(n+k)}\right\}=0.\label{fnm}
\end{align}

in which $0\leq k \leq \min\{n_s-m,n_a-n\}$.

 \emph{Proof of}\textbf{Property 2}.

 Eq.~(\ref{recurs}) can be rewritten in the following form:

\begin{multline}\label{recurs2}
\Q_{\psi_i}^{(m)\dag}\Q_{a}^{(n)}=\Q_{a}^{(n)}\Q_{\psi_i}^{(m)\dag}-\\-\sum_{k}
k!\binom{m+k}{k}\!\!\binom{n+k}{k}
\Q_{\psi_i}^{(m+k)\dag}\Q_{a}^{(n+k)}
\end{multline}

where $1\leq k \leq\min\{n_s-m,n_a-n\}$. Inserting
eq.~(\ref{recurs2}) in the definition (\ref{cij}) it can be shown
that:

\begin{align}\label{recurs1}
&{\cal C}_{n,m}^{(s)}(i,j)=\\&=\bra{0}\Q_a^{(n_a-s+n)\dag}
\Q_a^{(n_a-s+n)}\Q_{\psi_i}^{(n_s-n)\dag}\Q_{\psi_j}^{(n_s-n)}
\ket{0}-\nonumber\\&-\sum_{k}
k!\binom{n_a-s+m+k}{k}\!\!\binom{n_s-n+k}{k} {\cal
C}_{n-k,m}^{(s-k)}(i,j)\nonumber
\end{align}

with $1\leq k \leq\min\{n,s-m\}$.
 As all the states $\psi_i$ contain a fixed number $L$ of photons
$\tilde{P}_{\psi_i}(c^\dag,d_k^\dag)$ is a homogeneous polynomial
of degree $L$ in $c^\dag$ and $d_k^\dag$ and therefore the generic
$\Q_{\psi_i}^{(n)}$ is a homogeneous polynomial of degree $L-n$ in
$d_k^\dag$. As a consequence
$\Q_{\psi_i}^{(n_s-n)\dag}\Q_{\psi_j}^{(n_s-m)}\ket{0}=0$ unless
$n\leq m$. From this follows that for $n \geq m$ the first term on
the right hand side of eq.~(\ref{recurs1}) can be expressed as:
\begin{equation}\label{delta}
  \delta_{n,m}\bra{0}\Q_{\psi_i}^{(n_s-n)\dag}\Q_{\psi_j}^{(n_s-n)}\ket{0}\bra{0}
\Q_a^{(n_a-s+n)\dag} \Q_a^{(n_a-s+n)}\ket{0}
\end{equation}
Above we have introduced the completeness relation
$\sum_{\{\mathbf{n}\}}\ket{\mathbf{n}}\bra{\mathbf{n}}$, where
$\ket{\mathbf{n}}$ is a Fock states of the relevant modes. Note
that only the term corresponding to $\ket{0}\bra{0}$ survives.
Hence the proof is complete.

 \emph{Proof of} \textbf{Property1}.

By definition:
\begin{eqnarray}
 {\cal C}_{m,n}(i,j) &=&
 \bra{0}\Q_{\psi_i}^{(n_s-m)\dag}\Q_a^{(n_a-n)\dag}
 \Q_a^{(n_a-m)} \Q_{\psi_j}^{(n_s-n)}\ket{0}\nonumber\\
 &=&\left(\bra{0}\Q_{\psi_j}^{(n_s-n)\dag} \Q_a^{(n_a-m)\dag}
 \Q_a^{(n_a-n)}  \Q_{\psi_i}^{(n_s-m)}\ket{0}\right)^*\nonumber\\
& =& {\cal C}^*_{n,m}(j,i)
\end{eqnarray}
from the above equation and eq.(\ref{recurrence}) immediately
follows that for $m<n$

\begin{align}
&{\cal C}_{m,n}^{(s)}(i,j)=\\&-\sum_{k=1}^E
k!\binom{n_a-s+m+k}{k}\!\!\binom{n_s-n+k}{k} {\cal
C}_{m,n-k}^{(s-k)}(i,j)\nonumber
\end{align}

As ${\cal C}_{0,0}^{(0)}(i,j)$ is trivially symmetric, it follows
from recursion on $s$, that all ${\cal C}_{n,m}^{(s)}(i,j)$ are
symmetric under exchange of indices $n$ and $m$.

\end{document}